

\font\twelverm=cmr10 scaled 1200    \font\twelvei=cmmi10 scaled 1200
\font\twelvesy=cmsy10 scaled 1200   \font\twelveex=cmex10 scaled 1200
\font\twelvebf=cmbx10 scaled 1200   \font\twelvesl=cmsl10 scaled 1200
\font\twelvett=cmtt10 scaled 1200   \font\twelveit=cmti10 scaled 1200
\skewchar\twelvei='177   \skewchar\twelvesy='60
\def\twelvepoint{\normalbaselineskip=12.4pt
  \abovedisplayskip 12.4pt plus 3pt minus 9pt
  \belowdisplayskip 12.4pt plus 3pt minus 9pt
  \abovedisplayshortskip 0pt plus 3pt
  \belowdisplayshortskip 7.2pt plus 3pt minus 4pt
  \smallskipamount=3.6pt plus1.2pt minus1.2pt
  \medskipamount=7.2pt plus2.4pt minus2.4pt
  \bigskipamount=14.4pt plus4.8pt minus4.8pt
  \def\rm{\fam0\twelverm}          \def\it{\fam\itfam\twelveit}%
  \def\sl{\fam\slfam\twelvesl}     \def\bf{\fam\bffam\twelvebf}%
  \def\mit{\fam 1}                 \def\cal{\fam 2}%
  \def\tt{\twelvett}
  \textfont0=\twelverm   \scriptfont0=\tenrm   \scriptscriptfont0=\sevenrm
  \textfont1=\twelvei    \scriptfont1=\teni    \scriptscriptfont1=\seveni
  \textfont2=\twelvesy   \scriptfont2=\tensy   \scriptscriptfont2=\sevensy
  \textfont3=\twelveex   \scriptfont3=\twelveex  \scriptscriptfont3=\twelveex
  \textfont\itfam=\twelveit
  \textfont\slfam=\twelvesl
  \textfont\bffam=\twelvebf \scriptfont\bffam=\tenbf
  \scriptscriptfont\bffam=\sevenbf
  \normalbaselines\rm}

\def\beginlinemode{\endmode
  \begingroup\parskip=0pt \obeylines\def\\{\par}\def\endmode{\par\endgroup}}
\def\beginparmode{\endmode
  \begingroup \def\endmode{\par\endgroup}}
\let\endmode=\par
{\obeylines\gdef\
{}}
\def\singlespace{\baselineskip=\normalbaselineskip}
\def\oneandahalfspace{\baselineskip=\normalbaselineskip
  \multiply\baselineskip by 3 \divide\baselineskip by 2}
\def\doublespace{\baselineskip=\normalbaselineskip \multiply\baselineskip by 2}
\newcount\firstpageno
\firstpageno=2
\footline={\ifnum\pageno<\firstpageno{\hfil}\else{\hfil\twelverm\folio\hfil}%
\fi}
\let\rawfootnote=\footnote              
\def\footnote#1#2{{\rm\singlespace\parindent=0pt\rawfootnote{#1}{#2}}}
\def\raggedcenter{\leftskip=2em plus 12em \rightskip=\leftskip
  \parindent=0pt \parfillskip=0pt \spaceskip=.3333em \xspaceskip=.5em
  \pretolerance=9999 \tolerance=9999
  \hyphenpenalty=9999 \exhyphenpenalty=9999 }
\parskip=\medskipamount
\twelvepoint            
\overfullrule=0pt       
\def\preprintno#1{
 \rightline{\rm #1}}    
\def\author                     
  {\vskip 3pt plus 0.2fill \beginlinemode
   \singlespace \raggedcenter \twelvesc}
\def\affil                      
  {\vskip 3pt plus 0.1fill \beginlinemode
   \oneandahalfspace \raggedcenter \sl}
\def\abstract                   
  {\vskip 3pt plus 0.3fill \beginparmode
   \doublespace \narrower \noindent ABSTRACT: }
\def\endtitlepage               
  {\endpage                     
   \body}
\def\body                       
  {\beginparmode}               

\def\subhead#1{                 
  \vskip 0.1truein             
  {\raggedcenter #1 \par}
   \nobreak\vskip 0.1truein\nobreak}
\def\refto#1{$|{#1}$}           
\def\references                 
  {\subhead{References}         
   \beginparmode
   \frenchspacing \parindent=0pt \leftskip=1truecm
   \parskip=8pt plus 3pt \everypar{\hangindent=\parindent}}
\gdef\refis#1{\indent\hbox to 0pt{\hss#1.~}}    
\gdef\journal#1, #2, #3, 1#4#5#6{               
    {\sl #1~}{\bf #2}, #3, (1#4#5#6)}           
\def\refstylenp{                
  \gdef\refto##1{ [##1]}                                
  \gdef\refis##1{\indent\hbox to 0pt{\hss##1)~}}        
  \gdef\journal##1, ##2, ##3, ##4 {                     
     {\sl ##1~}{\bf ##2~}(##3) ##4 }}
\def\refstyleprnp{              
  \gdef\refto##1{ [##1]}                                
  \gdef\refis##1{\indent\hbox to 0pt{\hss##1)~}}        
  \gdef\journal##1, ##2, ##3, 1##4##5##6{               
    {\sl ##1~}{\bf ##2~}(1##4##5##6) ##3}}

\def\prd{\journal Phys. Rev. D, }
\def\prl{\journal Phys. Rev. Lett., }
\def\prpts{\journal Phys. Reports, }
\def\np{\journal Nucl. Phys., }
\def\pl{\journal Phys. Lett., }
\def\endreferences{\body}
\def\endpage                    
  {\vfill\eject}
\def\endpaper                   
  {\endmode\vfill\supereject}
\def\endit
  {\endpaper\end}
\def\ref#1{Ref. #1}                     
\def\Ref#1{Ref. #1}                     

\def\m@th{\mathsurround=0pt }
\font\twelvesc=cmcsc10 scaled 1200
\def\cite#1{{#1}}
\def\(#1){(\call{#1})}
\def\call#1{{#1}}
\def\taghead#1{}
\def\leaderfill{\leaders\hbox to 1em{\hss.\hss}\hfill}
\def\twiddle{\lower.9ex\rlap{$\kern-.1em\scriptstyle\sim$}}
\def\bigtwiddle{\lower1.ex\rlap{$\sim$}}
\def\gtwid{\mathrel{\raise.3ex\hbox{$>$\kern-.75em\lower1ex\hbox{$\sim$}}}}
\def\ltwid{\mathrel{\raise.3ex\hbox{$<$\kern-.75em\lower1ex\hbox{$\sim$}}}}
\def\square{\kern1pt\vbox{\hrule height 1.2pt\hbox{\vrule width 1.2pt\hskip 3pt
   \vbox{\vskip 6pt}\hskip 3pt\vrule width 0.6pt}\hrule height 0.6pt}\kern1pt}
\catcode`@=11
\newcount\tagnumber\tagnumber=0

\immediate\newwrite\eqnfile
\newif\if@qnfile\@qnfilefalse
\def\write@qn#1{}
\def\writenew@qn#1{}
\def\w@rnwrite#1{\write@qn{#1}\message{#1}}
\def\@rrwrite#1{\write@qn{#1}\errmessage{#1}}

\def\taghead#1{\gdef\t@ghead{#1}\global\tagnumber=0}
\def\t@ghead{}

\expandafter\def\csname @qnnum-3\endcsname
  {{\t@ghead\advance\tagnumber by -3\relax\number\tagnumber}}
\expandafter\def\csname @qnnum-2\endcsname
  {{\t@ghead\advance\tagnumber by -2\relax\number\tagnumber}}
\expandafter\def\csname @qnnum-1\endcsname
  {{\t@ghead\advance\tagnumber by -1\relax\number\tagnumber}}
\expandafter\def\csname @qnnum0\endcsname
  {\t@ghead\number\tagnumber}
\expandafter\def\csname @qnnum+1\endcsname
  {{\t@ghead\advance\tagnumber by 1\relax\number\tagnumber}}
\expandafter\def\csname @qnnum+2\endcsname
  {{\t@ghead\advance\tagnumber by 2\relax\number\tagnumber}}
\expandafter\def\csname @qnnum+3\endcsname
  {{\t@ghead\advance\tagnumber by 3\relax\number\tagnumber}}

\def\equationfile{%
  \@qnfiletrue\immediate\openout\eqnfile=\jobname.eqn%
  \def\write@qn##1{\if@qnfile\immediate\write\eqnfile{##1}\fi}
  \def\writenew@qn##1{\if@qnfile\immediate\write\eqnfile
    {\noexpand\tag{##1} = (\t@ghead\number\tagnumber)}\fi}
}

\def\callall#1{\xdef#1##1{#1{\noexpand\call{##1}}}}
\def\call#1{\each@rg\callr@nge{#1}}

\def\each@rg#1#2{{\let\thecsname=#1\expandafter\first@rg#2,\end,}}
\def\first@rg#1,{\thecsname{#1}\apply@rg}
\def\apply@rg#1,{\ifx\end#1\let\next=\relax%
\else,\thecsname{#1}\let\next=\apply@rg\fi\next}

\def\callr@nge#1{\calldor@nge#1-\end-}
\def\callr@ngeat#1\end-{#1}
\def\calldor@nge#1-#2-{\ifx\end#2\@qneatspace#1 %
  \else\calll@@p{#1}{#2}\callr@ngeat\fi}
\def\calll@@p#1#2{\ifnum#1>#2{\@rrwrite{Equation range #1-#2\space is bad.}
\errhelp{If you call a series of equations by the notation M-N, then M and
N must be integers, and N must be greater than or equal to M.}}\else%
 {\count0=#1\count1=#2\advance\count1
by1\relax\expandafter\@qncall\the\count0,%
  \loop\advance\count0 by1\relax%
    \ifnum\count0<\count1,\expandafter\@qncall\the\count0,%
  \repeat}\fi}

\def\@qneatspace#1#2 {\@qncall#1#2,}
\def\@qncall#1,{\ifunc@lled{#1}{\def\next{#1}\ifx\next\empty\else
  \w@rnwrite{Equation number \noexpand\(>>#1<<) has not been defined yet.}
  >>#1<<\fi}\else\csname @qnnum#1\endcsname\fi}

\let\eqnono=\eqno
\def\eqno(#1){\tag#1}
\def\tag#1$${\eqnono(\displayt@g#1 )$$}

\def\aligntag#1\endaligntag
  $${\gdef\tag##1\\{&(##1 )\cr}\eqalignno{#1\\}$$
  \gdef\tag##1$${\eqnono(\displayt@g##1 )$$}}

\def\eqalignno#1{\displ@y \tabskip\centering
  \halign to\displaywidth{\hfil$\displaystyle{##}$\tabskip\z@skip
    &$\displaystyle{{}##}$\hfil\tabskip\centering
    &\llap{$\displayt@gpar##$}\tabskip\z@skip\crcr
    #1\crcr}}

\def\displayt@gpar(#1){(\displayt@g#1 )}

\def\displayt@g#1 {\rm\ifunc@lled{#1}\global\advance\tagnumber by1
        {\def\next{#1}\ifx\next\empty\else\expandafter
        \xdef\csname @qnnum#1\endcsname{\t@ghead\number\tagnumber}\fi}%
  \writenew@qn{#1}\t@ghead\number\tagnumber\else
        {\edef\next{\t@ghead\number\tagnumber}%
        \expandafter\ifx\csname @qnnum#1\endcsname\next\else
        \w@rnwrite{Equation \noexpand\tag{#1} is a duplicate number.}\fi}%
  \csname @qnnum#1\endcsname\fi}

\def\ifunc@lled#1{\expandafter\ifx\csname @qnnum#1\endcsname\relax}

\let\@qnend=\end\gdef\end{\if@qnfile
\immediate\write16{Equation numbers written on []\jobname.EQN.}\fi\@qnend}

\catcode`@=12
\catcode`@=11
\newcount\r@fcount \r@fcount=0
\def\refreset{\global\r@fcount=0}
\newcount\r@fcurr
\immediate\newwrite\reffile
\newif\ifr@ffile\r@ffilefalse
\def\w@rnwrite#1{\ifr@ffile\immediate\write\reffile{#1}\fi\message{#1}}

\def\writer@f#1>>{}
\def\referencefile{
  \r@ffiletrue\immediate\openout\reffile=\jobname.ref%
  \def\writer@f##1>>{\ifr@ffile\immediate\write\reffile%
    {\noexpand\refis{##1} = \csname r@fnum##1\endcsname = %
     \expandafter\expandafter\expandafter\strip@t\expandafter%
     \meaning\csname r@ftext\csname r@fnum##1\endcsname\endcsname}\fi}%
  \def\strip@t##1>>{}}

\def\citeall#1{\xdef#1##1{#1{\noexpand\cite{##1}}}}
\def\cite#1{\each@rg\citer@nge{#1}}	

\def\each@rg#1#2{{\let\thecsname=#1\expandafter\first@rg#2,\end,}}
\def\first@rg#1,{\thecsname{#1}\apply@rg}	
\def\apply@rg#1,{\ifx\end#1\let\next=\relax
\else,\thecsname{#1}\let\next=\apply@rg\fi\next}

\def\citer@nge#1{\citedor@nge#1-\end-}	
\def\citer@ngeat#1\end-{#1}
\def\citedor@nge#1-#2-{\ifx\end#2\r@featspace#1 
  \else\citel@@p{#1}{#2}\citer@ngeat\fi}	
\def\citel@@p#1#2{\ifnum#1>#2{\errmessage{Reference range #1-#2\space is bad.}%
    \errhelp{If you cite a series of references by the notation M-N, then M and
    N must be integers, and N must be greater than or equal to M.}}\else%
 {\count0=#1\count1=#2\advance\count1
by1\relax\expandafter\r@fcite\the\count0,%
  \loop\advance\count0 by1\relax
    \ifnum\count0<\count1,\expandafter\r@fcite\the\count0,%
  \repeat}\fi}

\def\r@featspace#1#2 {\r@fcite#1#2,}	
\def\r@fcite#1,{\ifuncit@d{#1}
    \newr@f{#1}%
    \expandafter\gdef\csname r@ftext\number\r@fcount\endcsname%
                     {\message{Reference #1 to be supplied.}%
                      \writer@f#1>>#1 to be supplied.\par}%
 \fi%
 \csname r@fnum#1\endcsname}
\def\ifuncit@d#1{\expandafter\ifx\csname r@fnum#1\endcsname\relax}%
\def\newr@f#1{\global\advance\r@fcount by1%
    \expandafter\xdef\csname r@fnum#1\endcsname{\number\r@fcount}}

\let\r@fis=\refis			
\def\refis#1#2#3\par{\ifuncit@d{#1}
   \newr@f{#1}%
   \w@rnwrite{Reference #1=\number\r@fcount\space is not cited up to now.}\fi%
  \expandafter\gdef\csname r@ftext\csname r@fnum#1\endcsname\endcsname%
  {\writer@f#1>>#2#3\par}}

\def\ignoreuncited{
   \def\refis##1##2##3\par{\ifuncit@d{##1}%
     \else\expandafter\gdef\csname r@ftext\csname
r@fnum##1\endcsname\endcsname%
     {\writer@f##1>>##2##3\par}\fi}}

\def\r@ferr{\endreferences\errmessage{I was expecting to see
\noexpand\endreferences before now;  I have inserted it here.}}
\let\r@ferences=\references
\def\references{\r@ferences\def\endmode{\r@ferr\par\endgroup}}

\let\endr@ferences=\endreferences
\def\endreferences{\r@fcurr=0
  {\loop\ifnum\r@fcurr<\r@fcount
    \advance\r@fcurr by 1\relax\expandafter\r@fis\expandafter{\number\r@fcurr}%
    \csname r@ftext\number\r@fcurr\endcsname%
  \repeat}\gdef\r@ferr{}\global\r@fcount=0\endr@ferences}

\let\r@fend=\endpaper\gdef\endpaper{\ifr@ffile
\immediate\write16{Cross References written on []\jobname.REF.}\fi\r@fend}

\catcode`@=12

\citeall\refto		
\citeall\ref		%
\citeall\Ref		%

\referencefile
\def\uY{{U_1^{\ssc Y}}}
\def\uEM{{U_1^{\ssc EM}}}
\def\suL{{SU_2^{\ssc L}}}

\def\suc{{SU_3^{c}}}
\def\suPS{{SU_4^{\ssc PS}}}
\def\uBL{{U_1^{\ssc B-L}}}
\def\uX{{U_1^{\ssc X}}}
\def\uR{{U_1^{\ssc R}}}
\def\H{{\bar H}}
\def\ssc{\scriptscriptstyle}
\def\half{1/2}
\def\frac#1/#2{#1 / #2}
\def\ufgrant{This work was supported in part by the Institute for Fundamental
Theory and by DOE contract DE-FG05-86-ER40272.}
\def\uof{Department of Physics\\University of Florida\\Gainesville, FL 32611}
\def\oneandfourfifthsspace{\baselineskip=\normalbaselineskip
  \multiply\baselineskip by 9 \divide\baselineskip by 5}

\font\titlefont=cmr10 scaled\magstep3
\def\bigtitle                      
  {\null\vskip 3pt plus 0.2fill
   \beginlinemode \doublespace \raggedcenter \titlefont}

\oneandfourfifthsspace
\preprintno{UFIFT-HEP-92-22}
\preprintno{July, 1992}
\bigtitle{Some Simple Criteria for Gauged R-parity} \bigskip
\author Stephen P. Martin{$^\dagger$}
\affil\uof\body
\abstract
We catalog some simple conditions which are sufficient to guarantee that
R-parity survives as an unbroken gauged discrete subgroup of the continuous
gauge symmetry in certain supersymmetric extensions of the standard model.

\footnote{}{$^\dagger$Address after
Sept.~1, 1992: Dept.~of Physics, Northeastern University, Boston MA 02115}

\endtitlepage
\oneandfourfifthsspace

Low energy $N=1$ supersymmetry has been proposed as a cure for the fine-tuning
problem associated with the Higgs scalar boson[\cite{reviews}]. However,
in the minimal supersymmetric extension of the standard model, proton
decay might be expected to occur at an unacceptable rate due to the virtual
exchange of the superpartners of the standard model
states[\cite{rp},\cite{protondecay}].
To see this, we can write all of the renormalizable and gauge-invariant
terms which might occur in the superpotential:
$$
\eqalignno{
& W = W_1 + W_2 + W_3
&(sp1) \cr
& W_1 = \mu  H \H + y_u Q \H u + y_d Q H d + y_e LH e
&(sp2) \cr
& W_2 = \lambda_1 udd
&(sp3) \cr
& W_3 = \mu^\prime L \H +
\lambda_2 QLd +
\lambda_3 LLe
\>\>\> .
&(sp4) \cr
}
$$
[Here $Q$ and $L$ are chiral superfields for the $\suL$-doublet quarks and
leptons; $u$, $d$, $e$ are chiral superfields for the $\suL$-singlet quarks
and leptons, and $H$, $\H$ are the two $\suL$-doublet Higgs chiral superfields.
Family and gauge indices are suppressed. It is possible to eliminate
$\mu^\prime$ by a suitable rotation among the superfields $H$ and $L$;
but we choose not to do this because in most extensions of
the minimal supersymmetric standard model $H$ and $L$ will not
have the same quantum numbers.]
The terms in $W_1$ are just the supersymmetric versions of the usual standard
model Yukawa couplings and Higgs mass, and they conserve baryon number ($B$)
and lepton number ($L$). However, $W_2$ violates $B$ by one unit and $W_3$
violates $L$ by one unit. To prevent the proton from decaying in short order,
either $(\lambda_1)$ or $(\mu^\prime, \lambda_2, \lambda_3)$
must be very small[\cite{constraints}].

The simplest way to save the proton and avoid other phenomenological
disasters is to just banish all
of the terms occuring in $W_2$ and $W_3$ by means of a discrete $Z_2$ symmetry
known as R-parity[\cite{rp},\cite{protondecay}]. All of the standard model
states are taken to be even under R-parity and their superpartners are taken
to be odd. All interactions are required to have even R-parity. This means
that particles with odd R-parity are always produced in pairs, and that the
lightest particle with odd R-parity must be stable. At the level of the
chiral superfields, this may be implemented by assigning $R_p = -1$ to
$Q,L,u,d,e$ and $R_p = +1$ to $H, \H$.
(This $R_p$ is trivially related to R-parity by a factor
of $-1$ for fermions and is usually called matter parity.)
Then the terms in $W_2$ and $W_3$ are forbidden because they are $R_p$-odd,
while the terms in $W_1$ are $R_p$-even and allowed.
The $R_p$ symmetry also forbids some $B$ and $L$-violating operators of
dimension five and higher. Other discrete symmetries are
possible[\cite{BHR},\cite{IR},\cite{constraints}] which allow some terms
from either $W_2$ or $W_3$, as well as different
combinations of higher dimension operators violating $B$ and $L$.

At the level of the minimal supersymmetric standard model, the imposition
of $R_p$, or any other discrete symmetry designed to protect the proton from
decay, appears to be quite {\it ad hoc}. This is no longer necessarily
true if one considers extensions of the minimal theory in which the standard
model gauge group is embedded in a larger gauge symmetry, even if the
additional gauge symmetry is broken. Indeed, $R_p$ actually follows
automatically in certain theories with gauged $B-L$ (e.g. some supersymmetric
grand unified theories), and moreover can survive the spontaneous breakdown of
the continuous gauge invariance to the standard model gauge group. This will
occur if certain conditions, which seem to us to be surprisingly mild, are met
by the order parameters of the theory. For some reason, this seems to have
been underemphasized in the literature.
In this paper, we will catalog some of the simple criteria which are
sufficient to guarantee that $R_p$ is an unbroken discrete gauge symmetry
for various choices of the gauge group, by classifying the possible group
transformation properties of the order parameters of the
theory as ``safe" or ``unsafe" for $R_p$. For our purposes, it is most
convenient to note that for each chiral superfield,
$$
R_p = (-1)^{3(B-L)}
\>\>\> .
\eqno(rp)
$$
This strongly suggests that we obtain gauged $R_p$ as the discrete
remnant of a gauged $\uBL$. In fact, with the $\uBL$ assignments
$Q\sim\frac 1/3$;
$L\sim -1$; $u,d \sim -\frac 1/3$; $e\sim 1$; and $H,\H \sim 0$, it is clear
that unbroken $\uBL$ forbids each of the terms in $W_2$ and $W_3$.
To guarantee that $R_p$ should remain unbroken even after
$\uBL$ is broken, it is necessary and sufficient to require that all
Higgs vacuum expectation values (or other order parameters)
carry $3(B-L)$ charges which are even integers[\cite{FIQ},\cite{IR}].
Following the general arguments of Krauss and Wilczek[\cite{KW}], $\uBL$
then breaks down to a gauged  $Z_2$ subgroup which, in view of \(rp),
is nothing other than $R_p$. Unlike a global symmetry, such a gauged
discrete symmetry will be respected by Planck-scale physics[\cite{KW}].

The most natural setting for gauged $\uBL$ occurs in the Pati-Salam unification
of color and lepton number: $\suPS \supset \suc \times \uBL$.
Under the gauge group $ \suPS \times \suL \times \uR$, the standard model quark
and lepton superfields transform as  $Q,L \sim ({\bf 4}, {\bf 2}, 0)$ and
$d,e \sim ({\bf \overline 4}, {\bf 1}, \half)$ and
$u,N \sim ({\bf \overline 4}, {\bf 1}, -\half)$. [Here $N$ is the
superfield for a neutrino which transforms as a singlet under the
standard model gauge group.] With unbroken $\suPS$, the couplings $\lambda_1$,
$\lambda_2$, and $\lambda_3$ clearly vanish by gauge
invariance, since the $\suPS$ direct products
${\bf \overline 4} \times {\bf \overline 4} \times {\bf \overline 4}$
and ${\bf  4} \times {\bf  4} \times {\bf \overline 4}$ contain
no singlets[\cite{Slansky}]. Also, gauge invariance of the allowed Yukawa
couplings in $W_1$ requires that $H$ transforms as a linear combination of
$( {\bf 1}, {\bf 2}, -\half)$ and the color singlet part of
$( {\bf 15}, {\bf 2}, -\half)$, and that $\H$ transforms
as a linear combination of $( {\bf 1}, {\bf 2}, \half)$ and
the color singlet part of
$( {\bf 15}, {\bf 2}, \half)$. It then follows that $\mu^\prime$ vanishes
as well, because it is not allowed
by $\suPS$. So unbroken $\suPS$ prohibits the same terms in
$W_2$ and $W_3$ that $R_p$ does. This is hardly a surprise, since
$R_p$ is a discrete subgroup of $\uBL$ which is contained in $\suPS$.

Of course, $\suPS$ and $\uBL$ must be broken if we are to obtain the standard
model gauge group. To avoid breaking $R_p$ in the process, it is necessary
and sufficient that all of the order parameters
have even $\suPS$ quadrality, since we observe that
$$
SU_4^{\ssc PS} \> {\rm quadrality} \> = \> 3(B-L) \qquad [{\rm mod }\> 4]
\>\>\> .
\eqno(quadrality)
$$
Note that the standard model Higgs superfields $H$ and $\H$
have zero $\suPS$ quadrality and thus do not break $R_p$ when they
acquire vacuum expectation values. Since the order parameters must
also be color singlets, we  find that they may transform under $\suPS$ as
${\bf 1}, {\bf 10}, {\bf 15}, {\bf 35} \ldots$
and their conjugates, which we refer to as ``safe" reps. The order parameters
should not occur in ``unsafe" reps ${\bf 4}, {\bf 20^{\prime\prime}},
{\bf 36} \ldots$ if we want $R_p$ to survive. In particular, the $\suL$-singlet
order parameters may occur in the ``safe" reps
$( {\bf 1}, {\bf 1}, 0)$; $( {\bf 10}, {\bf 1}, 1)$; $( {\bf 15}, {\bf 1}, 0)$;
$( {\bf 35}, {\bf 1}, 2)\ldots$ of $ \suPS \times \suL \times \uR$
and their conjugates, but not in the ``unsafe" reps
$( {\bf 4}, {\bf 1}, \half)$;
$( {\bf 20^{\prime\prime}}, {\bf 1}, -\frac 3/2)$; $\ldots$ and their
conjugates. As long as we arrange for the theory to only have order parameters
corresponding to ``safe" reps, then $R_p$ remains unbroken.

Actually, $3(B-L)$ is always an integer multiple of 6 for safe order
parameters in $\suPS$,
since they must also be color singlets. Thus the surviving discrete subgroup
of $\uBL$ is a $Z_6$; however, a $Z_3$ subgroup of this is just the
discrete center of $\suc$, which is already taken into account.
So the remaining $Z_2 = R_p$ is what really counts. Also note that if
all order parameters in the theory had zero $\suPS$ quadrality, then because
of \(quadrality) we would be left with a $Z_4$ which contains $R_p$ as a
subgroup and eliminates certain operators of dimension $\geq 5$
which are allowed by $R_p$. However, such a situation is very unlikely,
since obtaining a realistic neutrino mass spectrum via the seesaw
mechanism[\cite{seesaw}] requires a Majorana mass term for $N$, which in turn
requires an order parameter with quadrality $2$.

The grand unified theory based on $SO_{10}$ contains $\suPS$ as a subgroup,
and so we may expect to obtain a nice criterion for this case also.
The standard model quark and lepton superfields all transform as components of
the 16-dimensional spinor rep of $SO_{10}$: $Q,L,d,e,u,N \sim {\bf 16}$.
Now the absence of the couplings $\lambda_1$,
$\lambda_2$, and $\lambda_3$ follows in the language of
unbroken $SO_{10}$ from the group theory fact
${\bf 16} \times {\bf 16} \times {\bf 16}
\not\supset {\bf  1}$. Furthermore, since ${\bf 16} \times {\bf 16} =
{\bf 10}_{\ssc S} + {\bf 120}_{\ssc A} + {\bf 126}_{\ssc S}$, it must be that
$H$ and $\H$ are each linear combinations of appropriate
components of ${\bf 10}$, ${\bf \overline {126}}$, and ${\bf 120}$
(which couples families antisymmetrically) in order to allow the
Yukawa couplings in $W_1$. Then from
${\bf 16} \times {\bf 10} \not\supset {\bf 1}$,
${\bf 16} \times {\bf 120} \not\supset {\bf 1}$,
and ${\bf 16} \times {\bf \overline {126}} \not\supset {\bf 1}$ it follows
that $\mu^\prime = 0$  also for unbroken $SO_{10}$.

What happens after $SO_{10}$ is broken? The survival or non-survival of
$R_p$ just depends on the $SO_{10}$ transformation properties of the order
parameters. The relevant property of the $SO_{10}$ reps is the ``congruency
class"[\cite{Slansky}] which is defined mod 4. In fact, we find
$$
SO_{10} \> {\rm congruency}\>{\rm class} \>  = \>
3(B-L) \qquad [{\rm mod }\> 2],
\eqno(congruency)
$$
so that ``safe" reps for order parameters in $SO_{10}$ are just those with
congruency class 0 or 2. The safe reps are
$\bf 10$, $\bf 45$, $\bf 54$, $\bf 120$, $\bf 126$, $\bf 210$, $\bf
210^\prime$,
$\bf 320\, \ldots$ and their conjugates. The ``unsafe" reps are
$\bf 16$, $\bf 144$, $\bf 560\, \ldots$ and their conjugates.
As long as only order parameters corresponding to even $SO_{10}$ congruency
classes are used to break $SO_{10}$, the resulting low energy gauge group
will inevitably include an unbroken $R_p$.
The prospect of removing the unsafe reps for Higgs vacuum expectation
values from the model builder's palette seems unlamentable; for
example[\cite{HRR}], $SO_{10}$ can be broken down to
$\suc \times \suL \times \uY \times R_p$ with realistic Yukawa couplings and
then to $\suc \times \uEM \times R_p$,  using only order parameters in the reps
${\bf 10}$, ${\bf 54}$ and ${\bf \overline{126}}$.
Order parameters in unsafe reps are incapable of giving any masses to the
standard model states (or $N$) anyway. Note that the safe reps of $\suPS$
are precisely the ones which are embedded in safe reps of $SO_{10}$, since
$\suPS$ quadrality $=SO_{10}$ congruency class [mod 2].

(While chiral superfields in large reps may ruin the asymptotic freedom
of the unified gauge coupling, this need not concern us. The order
parameters associated with the large reps may find their place only in
a phenomenological description, and may not actually correspond to vacuum
expectation values for fundamental fields. Also, a Landau
singularity in the unified gauge coupling is presumeably irrelevant if it
occurs at a distance scale shorter than the Planck length.)

Note that gauged $\uBL$ does not occur in a pure $SU_5$ grand unified theory.
In fact, with the standard $SU_5$ assignments $L,d \sim {\bf \overline 5}$ and
$Q,u,e \sim {\bf 10}$, it is clear  that unbroken $SU_5$ allows
$\lambda_1$, $\lambda_2$ and $\lambda_3$ equally, by looking at the standard
model content of the $SU_5$ fact ${\bf \overline 5}\times
 {\bf \overline 5} \times {\bf 10} \supset {\bf 1}$.
Furthermore, $H$ consists of some ${\bf \overline 5}$ and some
${\bf \overline {45}}$ and $\H$ consists of some ${\bf  5}$ and some
${\bf  45}$; so $\mu^\prime$ is also certainly allowed. There is
no reason for any of these couplings to vanish in pure $SU_5$, in sharp
contrast
to our other examples.

The case of supersymmetric ``flipped" $SU_5$[\cite{flipped}]
is quite different from pure $SU_5$ since it
contains gauged $\uBL$. Under the gauge group $SU_5 \times U_1^f$, the
standard model fermions transform as $Q,d,N \sim ({\bf 10},1)$;
$L,u \sim ({\bf \overline 5}, -3)$; and
$e \sim ({\bf 1}, 5)$. Naturally, $W_2$ and $W_3$ are absent as long as
$SU_5 \times U_1^f$ is unbroken, which can be seen as a consequence of the
$\uBL$ subgroup. When
$SU_5 \times U_1^f$ breaks, $R_p$ survives if the order parameters transform
as components of certain ``safe" reps of
$SU_5 \times U_1^f$. Since $U_1^f$ charge $= 3(B-L)$ mod 2, the
safe reps are just those which have even integer
$U_1^f$ charges in our normalization.
These include $({\bf 5},-2)$ and $({\bf \overline 5},2)$; the unsafe
reps include $({\bf 5},3)$ and $({\bf 10},1)$ and their conjugates.
One of the selling points of flipped $SU_5$ is that
the spontaneous symmetry breaking can be accomplished using only Higgs
fields in reps no larger than the $({\bf 10},1)$. However, this cannot
be accomplished if one insists on using only Higgs fields which are safe for
$R_p$. If unbroken $R_p$ exists in such models, it must come from an additional
structure (e.g. superstring theory).

The grand unified theory $E_6$ also contains $\uBL$ as a subgroup.
However, reps of $E_6$ cannot be classified as safe or unsafe for
$R_p$, because each irreducible rep contains components with both even and
odd values of $3(B-L)$. Since $R_p$ is an abelian discrete subgroup, it will
suffice for us to classify superfields and
possible order parameters in $E_6$ according to their transformation properties
under the subgroup $\suc \times \suL \times \uY \times \uBL \times \uX$.
(It does not concern us whether this subgroup is actually the unbroken gauge
group at any particular stage of symmetry breaking.)
The ${\bf 27}$ of $E_6$ transforms under this subgroup as
$({\bf 3},{\bf 2}, \frac 1/6, \frac 1/3, 1) +
({\bf 1},{\bf 2}, -\half, -1, 3) +
({\bf {\overline 3}},{\bf 1}, -\frac 2/3, -\frac 1/3, 2) +
({\bf {\overline 3}},{\bf 1}, \frac 1/3, -\frac 1/3, 2) +
({\bf 1},{\bf 1}, 1, 1, 0) +
({\bf 1},{\bf 1}, 0, 1, 0) +
({\bf {\overline 3}},{\bf 1}, \frac 1/3, \frac 2/3, - 4) +
({\bf 3},{\bf 1}, -\frac 1/3, -\frac 2/3, -2) +
({\bf 1},{\bf 2}, \half, 0, -3) +
({\bf 1},{\bf 2}, -\half, 0, -3) +
({\bf 1},{\bf 1}, 0, 0, 6) $.
This defines $\uX$, which may be
thought of as the extra $U_1$ which occurs in $E_6$ but not $SO_{10}$.
The first five terms may be identified with $Q$, $L$, $u$, $d$, and $e$
respectively. (We assume that each of the three standard model families are
embedded in the ${\bf 27}$ in the same way.)
It then follows that $H$ and $\H$ transform as
$({\bf 1},{\bf 2}, -\half, 0, -3)$ and  $({\bf 1},{\bf 2}, \half, 0, -3)$.
With these assignments,
the Yukawa terms in $W_1$ all transform as
$({\bf 1},{\bf 1}, 0, 0, 0)$ and the terms $udd$, $QLd$ and $LLe$ in
$W_2$ and $W_3$ each transform as $({\bf 1},{\bf 1}, 0, -1, 6)$. The term
$L\H$ in $W_3$ transforms as $({\bf 1},{\bf 1}, 0, -1, 0)$. Note that
an order parameter $({\bf 1},{\bf 1}, 0, 0, 6)$ is necessary so that the
non-standard-model particles
$({\bf {\overline 3}},{\bf 1}, \frac 1/3, \frac 2/3, - 4) +
({\bf 3},{\bf 1}, -\frac 1/3, -\frac 2/3, -2)$ can get masses,
and to allow $H\H$ in $W_1$.
We now classify as safe or unsafe the possible $\Delta I=0$
and $\Delta I= \half$ order parameters which occur in the smallest few
reps of $E_6$, namely the $\bf 27$, $\bf 78$, $\bf 351$, and
$\bf 351^\prime$. (Safe and unsafe reps in
${\bf \overline {27}}$, ${\bf \overline {351}}$ and
${\bf \overline {351^\prime}}$ can be found by conjugating the ones below.)

The $\Delta I=0$ order parameters for $E_6$ which are safe for unbroken
$R_p$ are: $({\bf 1},{\bf 1}, 0, 0, 0)$, which occurs three times in the
$\bf 78$ of $E_6$;
$({\bf 1},{\bf 1}, 0, 0, -12)$, which occurs once  in the $\bf 351^\prime$;
$({\bf 1},{\bf 1}, 0, 0, 6)$, which occurs once in the ${\bf 27}$,
twice in the $\bf 351$, and once  in the $\bf 351^\prime$;
and $({\bf 1},{\bf 1}, 0, -2, 0)$, which occurs once  in the $\bf 351^\prime$.
The unsafe $\Delta I=0$ order parameters for $E_6$ are:
$({\bf 1},{\bf 1}, 0, -1, 6)$ and $({\bf 1},{\bf 1}, 0, 1, -6)$, which each
occur once in the $\bf 78$; $({\bf 1},{\bf 1}, 0, 1, 0)$, which occurs once
in the ${\bf 27}$, twice in the $\bf 351$, and once  in the $\bf 351^\prime$;
and $({\bf 1},{\bf 1}, 0, -1, -6)$, which occurs once  in the $\bf 351$ and
once  in the $\bf 351^\prime$.

The safe $\Delta I=\half$ order parameters for $E_6$ transform according to:
$({\bf 1},{\bf 2}, \half, 0, -3)$ and $({\bf 1},{\bf 2}, -\half, 0, -3)$, which
each occur once in the $\bf 27$, three times in the ${\bf 351}$ and twice
in the ${\bf 351^\prime}$. The unsafe $\Delta I= \half$ reps include
$({\bf 1},{\bf 2}, \half, 1, 3)$ and
$({\bf 1},{\bf 2}, -\half, -1, -3)$ which each occur once in the $\bf 78$;
$({\bf 1},{\bf 2}, -\half, -1, 3)$ which occurs once in the ${\bf 27}$,
three times in the $\bf 351$, and twice in the $\bf 351^\prime$;
$({\bf 1},{\bf 2}, \half, 1, -9)$, $({\bf 1},{\bf 2}, \half, -1, 3)$ and
$({\bf 1},{\bf 2}, -\frac 3/2, -1, 3)$ which each occur once in the $\bf 351$
and once in the $\bf 351^\prime$.

Now, using safe $\Delta I=0$ order parameters listed above for
$\bf 27$, $\bf 351$ and $\bf 351^\prime$, $E_6$ can be broken
down to $\suc \times \suL \times \uY \times R_p$ and
the states other than $Q,L,u,d,e$ in the $\bf 27$ are all eligible to
obtain large masses. Also, safe order parameters
for $({\bf 1},{\bf 2}, \half, 0, -3)$ and  $({\bf 1},{\bf 2}, -\half, 0, -3)$
provide for standard model masses. So, the statement for $E_6$ is that
the safe reps for order parameters correspond to those which are
necessary anyway to break $E_6$ to the standard model with the correct mass
spectrum. Order parameters in the unsafe reps are not needed for anything
from this point of view, although demanding their absence would eliminate some
otherwise attractive symmetry breaking patterns.

Dimension-five operators which violate $B$ and $L$ can also contribute
to proton decay. The only such operators consistent with supersymmetry and
the standard model gauge group which are not already forbidden by $R_p$ are
$[uude]_F$, $[QQQL]_F$ and $[LL\H \H]_F$. The first two of these terms are
also allowed by any gauge invariance contained in $SO_{10}$,
in view of the standard model content of the group theory fact
${\bf 16} \times {\bf 16} \times {\bf 16} \times {\bf 16} \supset {\bf 1}
+ {\bf 1}$. However, all three terms (and $H \H$ in $W_1$) are
prohibited by unbroken $E_6$ because of ${\bf 27} \times {\bf 27} \times
{\bf 27} \times {\bf 27} \not\supset{\bf 1}$, etc.; this is a simple
consequence of $E_6$ triality[\cite{Slansky}], since $Q,L,u,d,e,H,\H$ all
must have triality 1. Or, it may be viewed as a consequence of $\uX$
and $\uBL$ conservation. The fact that these terms are forbidden by an
unbroken $E_6$ gauge group may provide a means of suppressing them even more
heavily than naive expectation in the  low energy theory, although the extent
to which this is true is model dependent. The order parameter
$({\bf 1},{\bf 1}, 0, 0, 12)$ breaks $\uX$ to a $Z_{12}$ without allowing the
aforementioned  terms. An order parameter
$({\bf 1},{\bf 1}, 0, 0, 6)$, which is presumeably necessary as
noted earlier, will further break $\uX$ to a $Z_{6}$ and allow
$[uude]_F$, $[QQQL]_F$ and $[ H \H]_F$.
This $Z_6$ is nothing new, because it is just a subgroup of
$Z_3^c \times Z_2^{\ssc L}$, where $Z_3^c$ is the center of
$\suc$ and $Z_2^{\ssc L}$ is the center of $\suL$. The order parameter
$({\bf 1},{\bf 1}, 0, -2, 0)$ breaks $\uBL$ to $R_p$ and allows $[LL\H \H]_F$.

If $R_p$-unsafe order parameters do occur in theories with gauged
$\uBL$, then $R_p$ is spontaneously broken. This is not necessarily
a disaster if the mass scales of the unsafe order parameters are
sufficiently small compared to the unification scale. One popular example
which has been explored in the literature[\cite{spon}] involves spontaneous
$R_p$ breaking due to an expectation value for the scalar partner
of a neutrino which transforms in the unsafe rep
$({\bf 1},{\bf 1},0,1,0) \subset {\bf 27}$ of
$\suc \times \suL \times \uY \times \uBL \times \uX \subset E_6$, or the
unsafe rep $\bf 16$ of $SO_{10}$. There are, of course, other possible
examples (of problematical phenomenological viability) for spontaneously
broken $R_p$ if other unsafe reps listed above obtain vacuum expectation
values.
The dominant contributions to $B$ and $L$-violating terms in the low
energy effective superpotential come from tree graphs with unsafe order
parameters on external legs. Note that explicit $R_p$
breaking is really a contradiction in terms for any model with gauged $\uBL$,
because of \(rp); it must be either exact or spontaneously broken.

In superstring-inspired models based on remnants of $E_6$, the existence of
$R_p$ depends on e.g.~the properties of the six-dimensional compactified
manifold. There are a bewildering plethora of possibilities for the vacuum,
some of which respect $R_p$ or other generalized matter parities, and some of
which do not. Note that it is not possible to break the $\uBL$ subgroup to
$R_p$ in these models {\it after} compactification, because the necessary
safe order parameter resides in the $\bf 351^\prime$ (or larger) rep
of $E_6$, and not in the $\bf 27$, $\bf\overline {27}$, or singlet reps
which are available for chiral superfields in the perturbative
field theory limit of superstring theory. In such models it is not possible
to understand in detail ``why" unbroken $R_p$ should exist without
understanding the compactification mechanism. In this paper we have
considered instead criteria which can be understood
using only an effective $N=1$ supersymmetric model. These criteria
are reassuringly not too restrictive, and are
especially crisp in the languages of $\uBL$, $\suPS$ and $SO_{10}$.
Hopefully, R-parity appears less {\it ad hoc} in this light.

I am grateful to Pierre Ramond for helpful comments.
\ufgrant

\references
\refis{Slansky}
All group theory conventions and facts used in this paper may be
found in R.~Slansky, \prpts 79, 1, 1981.

\refis{reviews}
For reviews, see H.~P.~Nilles,  \prpts 110, 1, 1984 and
H.~E.~Haber and G.~L.~Kane, \prpts 117, 75, 1985.

\refis{KW} L.~Krauss and  F.~Wilczek, \prl 62, 1221, 1989.

\refis{protondecay} S.~Weinberg, \prd 26, 287, 1982;
N.~Sakai and T.~Yanagida, \np B197, 83, 1982.

\refis{constraints} For precise constraints, see
references [1], [2], and [3] and
F.~Zwirner, \pl B132, 103, 1983;
L.~J.~Hall and M.~Suzuki, \np B231, 419, 1984;
I.~H.~Lee, \np B246, 120, 1984;
S.~Dawson, \np B261, 297, 1985;
R.~Barbieri and A.~Masiero, \np B267, 679, 1986;
S.~Dimopolous, L.~J.~Hall, \pl B207, 210, 1987;
V.~Barger, G.~F.~Giudice, and T.~Han, \prd 40, 2987, 1989;
S.~Dimopolous, R.~Esmailzadeh, L.~J.~Hall, J.-P.~Merlo, and
G.~D.~Starkman, \prd 41, 2099, 1990;
R.~Barbieri, D.~E.~Brahm, L.~J.~Hall and S.~D.~H.~Hsu, \pl B238, 86, 1990;
B.~A.~Campbell, S.~Davidson, J.~Ellis and K.~A.~Olive, \pl B256, 457, 1991;
H.~Dreiner and  G.~G.~Ross, \np B365, 597, 1991.

\refis{BHR}
M.~C.~Bento, L.~Hall and  G.~G.~Ross, \np B292, 400, 1987.

\refis{rp} G.~Farrar, P.~Fayet, \pl B76, 575, 1978;
S.~Dimopolous and H.~Georgi, \np B193, 150, 1981.

\refis{HRR} J.~A.~Harvey, P.~Ramond, and D.~B.~Reiss, \pl B92, 309, 1980;
\np B199, 223, 1982.

\refis{IR}  L.~E.~Ib\'a\~nez and  G.~G.~Ross, \pl B260, 291, 1991;
\np B368, 3, 1992.

\refis{seesaw} M.~Gell-Mann, P.~Ramond, and R.~Slansky, in Sanibel Talk,
CALT-68-709, Feb 1979, and in {\it Supergravity},
(North Holland, Amsterdam, 1979; T.~Yanagida, in {\it Proc. of the
Workshop on
Unified Theories and Baryon Number in the Universe}, Tsukuba, Japan, 1979,
edited by A. Sawada and A. Sugamoto (KEK Report No. 79-18, Tsukuba, 1979).

\refis{flipped}  A.~De R\'ujula, H.~Georgi, and S.~L.~Glashow,
\prl 45, 413, 1980;
S.~M.~Barr, \pl B112, 219, 1982;
I.~Antoniadis, J.~Ellis, J.~S.~Hagelin, and D.~V.~Nanopoulos,
\pl B199, 231, 1987.

\refis{FIQ}  A.~Font, L.~E.~Ib\'a\~nez and F.~Quevedo,
\pl B228, 79, 1989.

\refis{spon} C.~S.~Aulakh and R.~N.~Mohapatra,
\pl B119, 136, 1982;
J.~Ellis, G.~Gelmini, C.~Jarlskog, G.~G.~Ross, and J.~W.~F.~Valle,
\pl B150, 142, 1985;
G.~G.~Ross and J.~W.~F.~Valle,
\pl B151, 325, 1985;
B.~Gato, J.~Leon, J.~Perez-Mercader, and M.~Quiros,
\np B260, 203, 1985;
A.~Masiero and J.~W.~F.~Valle,
\pl B251, 273, 1990.

\endreferences\endit\end